\documentstyle[12pt,aasms4]{article}

\newcommand{\Lya}{\mbox{Ly$\alpha$}}
\newcommand{\hb}{\mbox{H$\beta$}}
\newcommand{\NV}{\mbox{N\,{\sc v}$\,\lambda$1240}}
\newcommand{\SiIV}{\mbox{Si\,{\sc iv}$\,\lambda$1400}}
\newcommand{\CIII}{\mbox{C\,{\sc iii}$\,\lambda$1909}}
\newcommand{\CIV}{\mbox{C\,{\sc iv}$\,\lambda$1549}}
\newcommand{\CIVnw}{\mbox{C\,{\sc iv}}}

\newcommand{\HeII}{\mbox{He\,{\sc ii}$\,\lambda$1640}}
\newcommand{\FeII}{\mbox{Fe\,{\sc ii}}}

\newcommand{\etal}{et al.\ }

\newcommand{\cfs}{\mbox{$F_{\lambda}$(1370\,\AA)}}
\newcommand{\cfl}{\mbox{$F_{\lambda}$(1855\,\AA)}}

\begin{document}

\title{
STEPS TOWARD DETERMINATION OF THE SIZE AND STRUCTURE \\
OF THE BROAD-LINE REGION IN ACTIVE GALACTIC NUCLEI. XIII. \\
ULTRAVIOLET OBSERVATIONS OF \\
THE BROAD-LINE RADIO GALAXY 3C~390.3}

\author{
P.T.~O'Brien,\altaffilmark{1} 
M.~Dietrich,\altaffilmark{2} 
K.~Leighly,\altaffilmark{3} 
D.~Alloin,\altaffilmark{4}  
J.~Clavel,\altaffilmark{5}  
D.M.~Crenshaw,\altaffilmark{6}  
R.A.~Edelson,\altaffilmark{1}
K.~Horne,\altaffilmark{7}  
G.A.~Kriss,\altaffilmark{8}  
J.H.~Krolik,\altaffilmark{8}
M.A.~Malkan,\altaffilmark{9}  
H.~Netzer,\altaffilmark{10}  
B.M.~Peterson,\altaffilmark{11} 
G.A.~Reichert,\altaffilmark{12}  
P.M.~Rodr\'iguez-Pascual,\altaffilmark{13}  
W.~Wamsteker,\altaffilmark{13}
K.S.J.~Anderson,\altaffilmark{14}  
N.G.~Bochkarev,\altaffilmark{15}
F.-Z.~Cheng,\altaffilmark{16}   
A.V.~Filippenko,\altaffilmark{17}
C.M.~Gaskell,\altaffilmark{18}
I.M.~George,\altaffilmark{19,20}
M.R.~Goad,\altaffilmark{21}
L.C.~Ho,\altaffilmark{22}
S.~Kaspi,\altaffilmark{10}
W.~Kollatschny,\altaffilmark{23}
K.T.~Korista,\altaffilmark{24}
G.~MacAlpine,\altaffilmark{25}
D.~Marlow,\altaffilmark{26}
P.G.~Martin,\altaffilmark{27}
S.L.~Morris,\altaffilmark{28}
R.W.~Pogge,\altaffilmark{11}
B.-C.~Qian,\altaffilmark{29}
M.C.~Recondo-Gonzalez,\altaffilmark{13}
J.M.~Rodr\'{\i}guez Espinosa,\altaffilmark{13}
M.~Santos-Lle\'o,\altaffilmark{30}
A.I.~Shapovalova,\altaffilmark{31}
J.M.~Shull,\altaffilmark{32}
G.M.~Stirpe,\altaffilmark{33}
W.-H.~Sun,\altaffilmark{34}
T.J.~Turner,\altaffilmark{19,20}
R.~Vio,\altaffilmark{13}
S.~Wagner,\altaffilmark{2}
I.~Wanders,\altaffilmark{11,7}
K.A.~Wills,\altaffilmark{26}
H.~Wu,\altaffilmark{35}
S.-J.~Xue,\altaffilmark{16}
and Z.-L.~Zou\altaffilmark{35}
}

\altaffiltext{1}{ Department of Physics and Astronomy, University of Leicester,
University Road, Leicester LE1\,7RH, United Kingdom.}

\altaffiltext{2}{ Landessternwarte, K\"{o}nigstuhl,
D-69117 Heidelberg, Germany.}

\altaffiltext{3}{ Columbia Astrophysics Laboratory, 538 West 120th Street,
New York, NY 10027.}

\altaffiltext{4}{ Centre d'Etudes de Saclay, Service d'Astrophysique,
Orme des Merisiers, 91191 Gif-Sur-Yvette Cedex, France.}

\altaffiltext{5}{ ISO Project, European Space Agency, Apartado 50727,
28080 Madrid, Spain.}

\altaffiltext{6}{ Computer Sciences Corporation, Laboratory for
Astronomy and Solar Physics, NASA Goddard Space Flight Center, Code 681,
Greenbelt, MD 20771.}

\altaffiltext{7}{ School of Physics and Astronomy, University of St.\ Andrews,
North Haugh, St.\ Andrews KY16\,9SS, Scotland, United Kingdom.}

\altaffiltext{8}{ Department of Physics and Astronomy, The Johns Hopkins
University, Baltimore, MD 21218.}

\altaffiltext{9}{ Department of Astronomy, University of California, 
Math-Science Building, Los Angeles, CA 90024.}

\altaffiltext{10}{ School of Physics and Astronomy and
the Wise Observatory, The Raymond and Beverly Sackler Faculty of Exact
Sciences, Tel-Aviv University, Tel-Aviv 69978, Israel.}

\altaffiltext{11}{ Department of Astronomy, The Ohio State University,
174 West 18th Avenue, Columbus, OH 43210.}

\altaffiltext{12}{ NASA Goddard Space Flight Center, Code 631,
Greenbelt, MD 20771.}

\altaffiltext{13}{ ESA IUE Observatory, P.O.\ Box 50727,
28080 Madrid, Spain.}

\altaffiltext{14}{ Department of Astronomy, New Mexico State University,
Box 30001, Dept.\ 4500, Las Cruces, NM 88003.}

\altaffiltext{15}{ Sternberg Astronomical Institute, M.V. Lomonosov
Moscow State University, Universitetskij Prospect, 13, Moscow 119899, Russia.}

\altaffiltext{16}{ Center for Astrophysics, University of Science
and Technology, Hefei, Anhui, People's Republic of China.}

\altaffiltext{17}{ Department of Astronomy, University of California,
Berkeley, CA 94720.}

\altaffiltext{18}{ Department of Physics and Astronomy,
University of Nebraska, Lincoln, NE 68588.}

\altaffiltext{19}{ Laboratory for High Energy Astrophysics,
NASA Goddard Space Flight Center, Greenbelt, MD 20771.}

\altaffiltext{20}{ Universities Space Research Association.}

\altaffiltext{21}{ Space Telescope Science Institute, 3700 San Martin Drive,
Baltimore, MD 21218.}

\altaffiltext{22}{ Harvard-Smithsonian Center for Astrophysics, 
60 Garden Street, Cambridge, MA 02138.}

\altaffiltext{23}{ Universit\"{a}ts-Sternwarte G\"{o}ttingen,
Geismarlandstrasse 11, D-37083 G\"{o}ttingen, Germany.}

\altaffiltext{24}{ Department of Physics and Astronomy,
University of Kentucky, Lexington, KY 40506.}

\altaffiltext{25}{ Department of Astronomy, University of Michigan,
Dennison Building, Ann Arbor, MI 48109.}

\altaffiltext{26}{ Nuffield Radio Astronomy Laboratory, University of 
Manchester, Jodrell Bank, Macclesfield, Cheshire, SK11 9DL, United Kingdom.}

\altaffiltext{27}{ Canadian Institute for Theoretical Astrophysics,
University of Toronto, ON, M5S 3H8 Canada.}

\altaffiltext{28}{ Dominion Astrophysical Observatory, 
5071 West Saanich Road, Victoria, B.C. V8X 4M6, Canada.}

\altaffiltext{29}{ Shanghai Astronomical Observatory,
Chinese Academy of Sciences, 80 Nandan Lu, Shanghai 200030, China.}

\altaffiltext{30}{ LAEFF, Apdo.\ 50727, E-28080 Madrid, Spain.}

\altaffiltext{31}{ Special Astrophysical Observatory,
Russian Academy of Sciences, Nizhni Arkhiz, Karachaj-Cherkess Region,
357147, Russia.}

\altaffiltext{32}{ Joint Institute for Laboratory Astrophysics,
University of Colorado, and National Institute of Standards and
Technology, Campus Box 440, Boulder, CO 80309.}

\altaffiltext{33}{ Osservatorio Astronomico di Bologna, Via Zamboni 33,
I-40126, Bologna, Italy.}

\altaffiltext{34}{ Institute of Astronomy, National Central University, 
Chung-Li, Taiwan 32054, Republic of China.}

\altaffiltext{35}{ Beijing Astronomical Observatory,
Chinese Academy of Sciences, Beijing 100080, People's Republic of China.}

\newpage

\begin{abstract}

As part of an extensive multi-wavelength monitoring campaign, the
International Ultraviolet Explorer satellite was used to observe the
broad-line radio galaxy 3C~390.3 during the period 1994 December 31 to
1996 March 5. Spectra were obtained every 6--10 days. The UV continuum
varied by a factor of 7 through the campaign, while the broad
emission-lines varied by factors of 2--5. Unlike previously monitored
Seyfert~1 galaxies, in which the X-ray continuum generally varies with a
larger amplitude than the UV, in 3C~390.3 the UV continuum light-curve is
similar in both amplitude and shape to the X-ray light-curve observed by
ROSAT. The UV broad emission-line variability lags that of the UV
continuum by 35--70 days for \Lya\ and \CIVnw ; values larger than those
found for Seyfert~1 galaxies of comparable UV luminosity. These lags are
also larger than those found for the Balmer lines in 3C~390.3 over the
same period. The red and blue wings of \CIVnw\ and \Lya\ vary in phase,
suggesting that radial motion does not dominate the kinematics of the UV
line-emitting gas. Comparison with archival data provides evidence for
velocity-dependent changes in the \Lya\ and \CIVnw\ line profiles,
indicating evolution in the detailed properties and/or distribution of
the broad-line emitting gas.

\end{abstract}


\section{Introduction}

The broad-line region (BLR) gas in active galactic nuclei (AGN) is
believed to be photoionized and heated by the AGN ionizing continuum
source, which in turn is generally believed to be powered by the release
of gravitational energy when matter is accreted onto a central massive
black hole. It is now well established that determining the variability
characteristics of AGN provides an invaluable tool with which to
investigate their physical nature (see \markcite{peterson93} Peterson
1993 and references therein). Indeed, AGN monitoring provides the only
practical method at present which can provide spatially resolved
information on the nuclear region immediately surrounding the black hole.
Broad emission-line variability in response to changes in the ionizing
continuum can be used to elucidate the BLR structure through the
technique of reverberation mapping (Blandford \& McKee 1982). The
multi-waveband continuum variability properties similarly constrain the
continuum emission mechanism.

The AGN ionizing continuum can vary significantly on timescales of a few
days or less and the light-crossing time of the BLR may be only a few
light-weeks in low-luminosity AGN. These two factors demand monitoring
campaigns with dense sampling patterns, while the unpredictable nature of
the continuum variability requires campaigns which typically last several
months in order to ensure a high probability of observing variations
suitable for measuring the line response. The International AGN Watch was
formed to overcome these practical difficulties and obtain data able to
place firm constraints on both the BLR structure and the origin of the
luminous AGN continuum. So far monitoring campaigns have been
successfully completed on five Seyfert~1 galaxies: NGC~5548 (Clavel \etal
1991; Peterson \etal 1991, 1992, 1994; Dietrich
\etal 1993; Maoz \etal 1993;
Korista \etal 1995; Romanishin \etal 1995; Marshall \etal 1997); 
NGC~3783 (Reichert \etal 1994; Stirpe \etal 1994; Alloin \etal 1995); 
NGC~4151 (Crenshaw \etal 1996; Kaspi \etal 1996; Warwick \etal 1996; 
Edelson \etal 1996); 
Fairall~9 (Rodr\'{\i}guez-Pascual \etal 1997; Santos-Lle\'o \etal 1997); and
NGC~7469 (Wanders \etal 1997; Nandra \etal 1998; Collier \etal 1998).

There are three principal observational results from the AGN Watch
campaigns which appear generic to those Seyfert~1 galaxies intensively
monitored: (1) The broad emission lines respond to continuum variations
with time-delays (lags) that range from a few days to months. (2) What
are commonly referred to as the ``high ionization'' emission lines (e.g.,
\Lya , \CIV ) respond on shorter timescales than the ``low ionization'' lines
(e.g., \hb, \CIII ), indicating ionization stratification of the BLR. (3)
Usually no statistically significant lag has been detected between the
optical and UV continuum variations. The only exception to this is
NGC~7469, for which Wanders \etal (1997) find a lag of $\sim 0.3$ days
between 1315\,\AA\ and 1825\,\AA\ using very densely sampled data. For
NGC~4151 the lag between the optical and UV continua and between the UV
and 1.5~keV X-ray continua are $\le 1$ and $\le 0.3$ days respectively
(Edelson \etal 1996). The lack of continuum variability time-delays is
perhaps the most fundamental result, being difficult to explain using
standard accretion disk models in which the variability signal travels
radially through the disk. This problem has supported development of
thermal reprocessing models in which X-ray photons are reprocessed into
UV and optical photons by an accretion disk or Thomson thick clouds
(Guilbert \& Rees 1988; Lightman \& White 1988). Although the details are
unclear, these models are capable of explaining at least some of the
observations (Krolik \etal 1991; Warwick
\etal 1996; Edelson \etal 1996).

The primary reason for the success of these monitoring programs has been
the unique capability of IUE to obtain well-calibrated UV spectra at
regular, closely-spaced intervals. However, despite the past success, it
must be stressed that the results summarized above have been obtained
only for low-luminosity, radio-quiet AGN. Coordinated UV/X-ray
monitoring has also only been possible for relatively short periods. To
fully test AGN models, it is essential that multi-wavelength data be
obtained for a wider range of AGN types over longer temporal baselines.
As part of this process, we recently carried out a large X-ray, UV,
optical, and radio monitoring campaign on the broad-line radio galaxy
(BLRG) 3C~390.3. In this paper we report on the UV campaign. The X-ray,
optical, and radio observations are presented elsewhere (Leighly \etal
1997; Dietrich \etal 1998).

The BLRG 3C~390.3 is an ideal target for monitoring. Optical continuum
variations of 30\% to 70\% on timescales of weeks to months are common
(e.g., Barr \etal 1980), together with large amplitude variations in both the
flux and profile of the Balmer emission lines (e.g., Veilleux \& Zheng
1992). Even larger amplitude continuum and broad-line variations on
similar timescales are seen in the under-sampled IUE archival data
(Clavel \& Wamsteker 1987; Zheng 1996; Wamsteker \etal 1997). The
redshift of 3C~390.3 ($z = 0.056$) is higher than that of the previously
monitored Seyfert~1 galaxies, resulting in a clean separation of the full
extent of the broad \Lya+\NV\ blend from the geocoronal \Lya\ line. Of
course the higher redshift is also a disadvantage in that the observed
flux is smaller, leading to lower signal-to-noise ratios than in our
previous campaigns. 

The complex broad-line profiles in 3C~390.3 suggest a highly structured
gas distribution and/or anisotropic illumination in the BLR. The Balmer
lines have localized peaks which are shifted to substantially higher and
lower velocities than the narrow lines. Evidence for such structure is
also seen in the \CIV\ and \Lya\ lines. This phenomenon of detached line
peaks is generally limited to BLRGs. Several explanations have been
proposed which the 3C~390.3 line profiles are particularly well suited to
test as they show unusually complex variations. One possibility is a
binary black hole system, each with an associated BLR (Gaskell 1988a;
Gaskell 1996). Eracleous \etal (1997) reject this hypothesis for
3C~390.3 as it implies black hole masses too high to reconcile with the
observed spectral and variability properties. An outflowing bi-conical
gas stream has also been proposed (Zheng \etal 1991), although if an
accretion disk is present, as seems essential to fuel the black hole, the
far-cone would most likely be obscured by the optically-thick disk (Livio
\& Xu 1997). Emission from an accretion disk (e.g., Perez
\etal 1988; Chen, Halpern, \& Filippenko 1989)
appears a more likely explanation. Rokaki, Boisson, \& Collin-Souffrin
(1992) find an accretion disk inclined at 30$^{\circ}$ fits both the
overall spectral-energy distribution and the broad-line profiles.
However, the Balmer line red and blue wings appear to vary independently
on long timescales (Zheng \etal 1991), in conflict with the simplest disk
models. The UV lines also show some evidence for this behaviour, although
analysis of the same IUE archival data have led to conflicting reports in
the literature. Wamsteker \etal (1997) find that the \Lya\ and \CIVnw\
blue wings lag the red, implying radial gas motion, but this result is not
confirmed by Zheng (1996). We note that the temporal sampling rate of the
archival data is low, with a median sampling interval of 94 days.

3C~390.3 is an edge-brightened radio double (Fanaroff-Riley type II
galaxy), has a relatively strong compact core, and displays superluminal
motion (Alef \etal 1996). The latter two properties are usually
interpreted as indicating a relativistic jet axis aligned close to the
line of sight (Orr \& Browne 1982). It is also a highly variable X-ray
source, and the X-ray spectrum shows a broad iron K$\alpha$ line (Inda
\etal 1994; Eracleous, Halpern, \& Livio 1996; Leighly \etal 1997).
Eracleous \etal (1996) find that an accretion disk inclined at
26$^{\circ}$ fits both the radio superluminal data and the X-ray spectra,
an inclination angle consistent with that derived by Rokaki \etal (1992).
Therefore, in addition to an accretion disk, some of the observed
ionizing continuum in 3C~390.3 may arise from a relativistic jet, unlike
that of the previously monitored radio-quiet AGN.

Relative to the Seyfert~1 galaxies previously monitored by AGN Watch,
3C~390.3 has a weak ``big blue bump''. Although the UV luminosity at
1370\AA , $L_{\lambda} \approx 1-2 \times 10^{40}$ erg s$^{-1}$
\AA$^{-1}$, is similar to that of NGC~5548, 3C~390.3 is a significantly
more luminous X-ray source. The weakness of the big blue bump 
and the historical X-ray light curve led Inda \etal (1994) to suggest
that some of the X-ray emission may arise from a similar process to that in
radio-loud QSOs, connected with the relativistic jet. Eracleous \etal
(1996) note that the X-ray variability in 3C~390.3 does not appear to be
related to the superluminal events, although the temporal sampling rate
of the published data is too poor to exclude a connection. The monitoring
campaign described here was designed to greatly improve on the temporal
sampling rates previously achieved so as to place firm constraints on the
BLR structure and continuum emission mechanisms in 3C~390.3.

\section{Observations}

The IUE SWP camera was used to obtain 65 low-dispersion (resolution
$\approx 6$\AA) observations of 3C~390.3 at typical intervals of 6--10
days from 1994 December 31 to 1996 March 5. Unfortunately the IUE campaign
had to be curtailed earlier than scheduled due to the loss of a gyro. The
observations were all made through the large ($10^{\prime\prime} \times
20^{\prime\prime}$) aperture. As 3C~390.3 is near the ecliptic pole, the
high angle between the telescope axis and the anti-solar direction (the
$\beta$ angle) meant that scattered solar light was present in the
telescope tube for most of the observations. This scattered light does
not contaminate the extracted SWP spectra (Weinstein \& Carini 1992), but
does raise the background in the Fine Error Sensor (FES) normally used to
maintain pointing accuracy during long, continuous exposures. For the
epochs when scattered light was present a number of short-exposure
segments, each typically 20 or 30 minutes long, were added together to form
each observation. The high $\beta$ angle also required the segmentation
of a few exposures to prevent over-heating the on-board computer which
could have led to data corruption. For the segmented observations the IUE
pointing accuracy was verified between each segment using the FES X-Y
position of a nearby bright offset star. This is a standard procedure
with IUE (Rodr\'{\i}guez-Pascual \etal\ 1997), and no evidence was found
for any significant guiding errors.

The only real consequences of the scattered light problem are a small loss in
on-target exposure time and the absence of an optical flux measurement
for the target from the FES. Details of the segmented exposures and FES
data can be found in the hand-written logs available at the IUE
observatories (observing program IDs SQ037, AGRMM and RQ079). The SWP
observation log is given in Table~1, in which the exposure time is the
total integration time from all the segments while the Julian Date
corresponds to the mid-point between the start of the first segment and
the end of the last. The UT date and time correspond to the start of
the first exposure segment.

\section{Data Reduction and Analysis}

The IUE spectral images were processed using the NEWSIPS pipeline
processing system (Nichols \etal 1993). This software incorporates a new
flat-fielding technique and an optimal extraction routine which together
greatly reduce the impact of detector fixed-pattern-noise and improve the
photometric accuracy. NEWSIPS also incorporates an improved absolute flux
calibration based on white-dwarf observations. The NEWSIPS output
products are FITS files, which contain many quality-control flags in the
header and comments from the hand-written logs such as details of the
exposure segments. The optimal extraction routine removes most cosmic ray
hits automatically. Together with spurious features due to imperfect
removal of reseaux marks, those few cosmic-rays which remained in the
extracted spectra (the MXLO files) were interpolated over before
measurement.

The quality-control flags of the NEWSIPS extraction routine allow a further
check on the location of the spectrum relative to the center of the large
aperture in the spatial direction. No spectrum was found to be more than
4 pixels away from the expected location, which is within the acceptable
range to prevent significant vignetting (Rodr\'{\i}guez-Pascual \etal
1997). The spectral location in the dispersion direction was similarly
checked by cross-correlating the narrow peaks of the \Lya\ and \CIV\
emission lines. The maximum shift found was only 2 pixels which is also
small enough to have caused no significant flux loss. Integer pixel
shifts were determined from the cross-correlation and applied to register the
spectra before further analysis. The broad-line peaks in the resulting
spectra give a redshift of $0.056\pm 0.001$, in excellent agreement with
that derived from the narrow optical lines (Eracleous \etal 1996).

As well as global quality flags, NEWSIPS provides quality flags at each
wavelength in the extracted spectra (the $\nu$ flag spectrum) which are
determined by an algorithm based on the percentage of pixels contributing
at each wavelength which suffer from a particular problem (Nichols \etal
1993). During the last third of the 3C~390.3 IUE campaign the
systematically higher continuum and emission-line fluxes led to most
spectra having a few pixels in the core of \Lya\ being flagged as
saturated. Where only a few pixels are saturated this is not a
significant problem. However, for seven of the spectra listed in Table~1
between five and ten pixels were flagged. Examination of the spectral
images, the $\nu$ flag spectrum, and comparison of these flagged spectra
with adjacent unsaturated spectra strongly suggest that all of the saturation
assignments by NEWSIPS for the 3C~390.3 spectra are borderline cases,
where the flux is incorrect by small amounts. Therefore these seven
spectra were included in the subsequent analysis, but the analysis was
repeated excluding them to check whether they significantly influenced
the results.

Of the 65 spectra listed in Table~1, five were excluded from the
analysis. Of these, four have exposure times less than 160 minutes and
are under-exposed. The other, SWP53825, has the unfortunate combination
of an unusually high background while the object was faint resulting in a
very noisy spectrum. The mean and median sampling intervals for the 60
good SWP spectra used in the subsequent analysis are 7.2 and 6.4 days
respectively.

\subsection{Measurements}

The average spectrum and the root-mean-square (rms) deviation from the
average derived from the 60 good SWP spectra are shown in Fig.~1. The
narrow-line components are virtually absent from the rms spectrum, which
is as expected since they are not predicted to vary significantly over
the timescale of this monitoring campaign. The rms spectrum is a measure
of the variable component of the UV spectrum and clearly shows that both
the broad emission lines and continuum varied during the campaign,
although the amplitude of variability was higher for the continuum.

All continuum and emission-line fluxes were measured in the observer's
frame. It is virtually impossible to find completely line-free continuum
wavebands in the UV, particularly longward of \CIV\ due to \FeII\ blends. The
unweighted average continuum flux was determined in two (observed)
wavebands: 1340--1400\,\AA\ and 1800--1910\,\AA , henceforth referred to as
``\cfs'' and ``\cfl'' respectively. For 3 spectra, noted in Table 2, the
\cfl\ waveband was truncated to avoid spurious features in the spectrum.
The continuum fluxes are given in Table 2 and plotted in Fig.~2.

The emission-line fluxes for each spectrum were determined by
integrating the flux above a first-order polynomial fitted to the
individual data within the continuum wavebands. For each line
fixed (observed) wavelength limits were used for the integration. For
\Lya\ and \CIV\ these limits were 1253.5--1313.5\,\AA\ and 1598.5--1674.5\,\AA\
respectively, corresponding to a velocity range of $\pm 7000$ km s$^{-1}$
relative to the (observed) line peaks at 1283.5\,\AA\ and 1636.5\,\AA . These
velocity limits correspond to those used by Wamsteker
\etal (1997). The \Lya\ and \CIV\ red wings will contain contributions
from \NV\ and \HeII\ (see below), but no attempt was made to deblend
these features. The total (blended) line fluxes will henceforth be
referred to as F(\Lya) and F(\CIVnw). To search for velocity-dependent
variations in line flux, the \Lya\ and \CIV\ lines were also measured
using three components --- red, center, and blue. The center corresponds to
$\pm 2000$ km s$^{-1}$ while the red and blue are $\pm$ (2000--7000) km
s$^{-1}$ respectively. The \Lya\ and \CIV\ fluxes are given in Table~3
and plotted in Fig.~3.

Determination of the emission-line fluxes for the weaker \NV , \SiIV , and
\HeII\ lines is more difficult due to the mediocre spectral quality. Based
on the rms spectra, the \NV\ flux was measured in the range 1310--1330\,\AA .
This overlaps with the \Lya\ measurement and only covers the red wing of
\NV\ to approximately +5000 km s$^{-1}$. For \SiIV\ limits of
1444--1512\,\AA\ were used corresponding to $\pm 7000$ km s$^{-1}$. For
\HeII\ the rms spectrum shows little evidence for variability, so the
flux was measured using limits of 1675--1790\,\AA\ ($\pm 10000$ km
s$^{-1}$). The \NV , \SiIV , and \HeII\ fluxes are given in Table~4 and
plotted in Fig.~4.

An additional ``\CIVnw\,(Far-red)'' flux was measured in the range
1670--1697\,\AA\ to study the possible fast variability of a ``shelf'' of
emission seen in the rms spectrum (Section~3.3.2) centered at $\approx
8500$ km s$^{-1}$. This feature is more prominent than in the rms
spectrum calculated by Wamsteker \etal (1997) from the IUE archival data.
A similar feature is also seen in the Balmer lines during 1994--95
(Dietrich \etal 1998).

The total and center emission-line fluxes include contributions from the
narrow-line region and are uncorrected for absorption either intrinsic to
3C~390.3 or due to our Galaxy. Spectra taken with the Faint Object
Spectrograph on the Hubble Space Telescope (Eracleous \etal 1998)
suggest that the narrow-line region contributes $\sim 20$\% of the mean
flux of the center component for \Lya\ and \CIV .

To estimate the uncertainties on the continuum and emission-line fluxes
we adopted the following procedure. First, ``observed'' flux
uncertainties were assigned at each epoch based on the spectral data
contributing to the continuum and emission-line fluxes. We then assume
there is no intrinsic variability between consecutive epochs $i, j$
within some monitoring period. The rms of the distribution of flux
differences (F$_j -$ F$_i$) for all the epochs within the monitoring
period is then calculated and compared to the mean observed uncertainty.
This provides a scale-factor with which to rescale all of the observed
uncertainties so that the rms of the flux differences is consistent with
that expected from the mean (rescaled) flux uncertainty. 

Rodr\'{\i}guez-Pascual \etal (1997) and Wanders \etal (1997) used an
analogous procedure to calculate the flux uncertainties for Fairall~9 and
NGC~7469, based on the rms of the distribution of flux ratios
(F$_j$/F$_i$) rather than flux differences. Applied to the 3C~390.3 data,
their procedure gives rescaled uncertainties for \Lya\ and
\CIVnw\ slightly larger than those used here, but gives
uncertainties 20--80\% larger for the weaker lines. Either procedure
should be considered as a conservative approach to calculating flux
uncertainties, as any intrinsic variability between adjacent epochs will
tend to increase the rms.

We used the entire light-curves to calculate the scale-factors, with
three exceptions --- \cfs , \cfl , and \CIVnw\,(Far-red). The relatively
large mean sampling interval for 3C~390.3 makes it likely that any
rapidly varying spectral component will have varied between observations.
This is likely for the UV continuum and the cross-correlation results
suggest it is also likely for \CIVnw\,(Far-red). Therefore, for these
three spectral components we rescaled using only those epochs within the
monitoring period from JD 2449810 -- 2449890. During this period both the
UV and X-ray continua (Leighly \etal 1997) were relatively quiescent.
Using the entire light-curves would increase the mean uncertainties for
\cfs\ and \cfl\ by 50 and 100\% respectively.

\subsection{Variability Analysis}

Various parameters characterising the continuum and emission-line
variability are given in Table~5. The mean flux, $\overline{F}$, and
standard deviation, $\sigma_F$, have their usual meanings. The $R_{max}$
parameter is the ratio of the maximum to minimum flux, and is only useful
when $\sigma_F$ is large compared to the measurement uncertainties. 
The values of $R_{max}$ are particularly uncertain for \SiIV\ and \HeII .
For \CIVnw\,(Far-red) the very high $R_{max}$ is due to the very low
flux at the first epoch. If that point is ignored $R_{max} = 7.24$.

The fourth parameter, $F_{var}$, in Table~5 is a measurement-uncertainty
corrected estimate of the amplitude of the variability relative to the
mean flux, defined as
\begin{equation}
F_{var} = {\sqrt{(\sigma_F^2 - \Delta^2)}\over{\overline{F}}},
\end{equation}
where $\Delta^2$ is the mean square value of the individual measurement
uncertainties, $\epsilon_i$, for the $i = 1,2,3, \ldots ,N$ observations , i.e.,
\begin{equation}
\Delta^2 = {1\over{N}} \sum_{i=1}^N \epsilon_i^2 .
\end{equation}

Both UV continuum light-curves show several distinct features which are also
seen, although more clearly defined, in the ROSAT X-ray light-curve of
3C~390.3 (Leighly \etal 1997). The UV continuum rises at the beginning of
the campaign, followed by a flare at JD~2449795 which coincides with a
large X-ray flare. The X-ray flare lasted about 15 days, so if the UV
flare were of comparable length we would expect only a single point above
the general trend in the data, as observed. After the flare the continuum
shows marginal evidence for flickering (the ``quiescent period''),
followed by a dip which is more pronounced at shorter wavelengths. The UV
continuum then rises to a maximum around JD~2450020, followed by a small
decline and then another rise at the end of the monitoring campaign.

The amplitude of the continuum variability is only slightly lower in the
longer wavelength band. In Fig.~5 we plot \cfl\ versus \cfs\ along with a
straight-line fit which allows for errors in both variables. As found for
Fairall~9 (Rodr\'iguez-Pascual \etal 1997), the straight-line fit is very
good ($\chi_{\nu}^2 = 1.4$), indicating that there is no significant
change in UV continuum shape with luminosity. The small positive
y-intercept may indicate a contribution to \cfl\ from weak blended \FeII\
emission lines and Balmer continuum of $(0.5\pm0.2) \times 10^{-15}$ erg
cm$^{-2}$ s$^{-1}$ \AA$^{-1}$. This small value (10\% of the mean \cfl )
and the absence of detectable \FeII\ in the rms spectrum (Fig.~1) show
that the \cfl\ variability parameters have not been significantly
affected by constant of weakly varying emission-line contamination.
Fitting to the logarithm of the continuum fluxes gives
\begin{equation}
\log \cfl = (-0.15\pm0.20) + (1.02\pm0.26) \log \cfs,
\end{equation}
consistent with a constant UV continuum shape ($f_{\nu} \propto
{\nu}^{-1}$) during the campaign.

All of the emission-line light-curves show a similar trend to that of the
continuum, with an overall increase in flux through the campaign. The
smallest $F_{var}$ amongst the emission lines occurs for \Lya, while
the others lines vary with similar amplitudes. Correcting
for the narrow-line components (Section 3.1) would increase $F_{var}$ by
$\sim 0.04$ for the \Lya\ and \CIVnw\ line centers, making them
comparable to those derived for the blue wings, but lower than for the
red wings. Unlike the \CIVnw\ line wings, $F_{var}$ for \Lya\,(Blue) is
almost half that for \Lya\,(Red).

Wamsteker \etal (1997) found that the \CIVnw/\Lya\ ratio for both the
blue and red wings increased with increasing continuum and \CIVnw\ flux
in the IUE archival data, but appeared to ``saturate'' for F(\CIVnw\,Red)
or F(\CIVnw\,Blue) $> 5 \times 10^{-13}$ erg cm$^{-2}$ s$^{-1}$. In
Fig.~6 we plot the \CIVnw/\Lya\ ratio correlations for both line wings
using the data from Table~3. Correlating against the line flux minimizes
the effect of time-delays in the line response which otherwise increase
the scatter. All four correlations shown in Fig.~6 are significant, with
Spearman rank correlation coefficients $>0.993$. 

As for the variability amplitudes, there is some evidence for contrasting
behaviour in the \CIVnw/\Lya\ ratio between the red and blue wings. For
\CIVnw\ line-wing fluxes $> 3 \times 10^{-13}$ erg cm$^{-2}$ s$^{-1}$ the
\CIVnw/\Lya\ ratio for the red wing falls systematically at the lower end
of the range for the blue wing (top boxes in Fig.~6) whereas the data for
both wings overlap at lower flux levels. Although this gives some support
to the idea of saturation at high fluxes levels in the red wing, the
trend is not as apparent as for the archival data. A difference between
the blue and red wings is more apparent from the correlation with
continuum flux, for which the \CIVnw/\Lya\ ratio is systematically higher
in the blue wing than the red (bottom boxes in Fig.~6). This trend is not
seen in the archival data. Overall, compared to either the continuum flux
or the line-wing flux, the line fluxes are lower and the average
\CIVnw/\Lya\ ratio is higher than found for the archival data. This is
particularly true for the blue wing for which the mean ratio is 50\%
higher. These results suggest a change in the behaviour of the
line-emitting gas. We will return to this point in the discussion.

\subsection{Cross-Correlation Analysis}

The time-delay or lag of the emission-line variation relative to
variations in the ionizing continuum is usually quantified by use of
a cross-correlation analysis. We performed two such analyses, using the
interpolation cross-correlation function (ICCF) as implemented by White
\& Peterson (1994) and their implementation of the discrete
cross-correlation function (DCF) of Edelson \& Krolik (1988). The CCF
results are tabulated in Table~6 and shown in Fig.~7. The ICCF and DCF
results are identical to within the DCF error-bars. Results which
are particularly uncertain are given within brackets in Table~6.

Due to the ``flat-topped'' nature of many of the CCFs for 3C~390.3 and
the relatively large flux uncertainties, the position of the CCF peak,
$\tau_{peak}$, is highly uncertain. Therefore, for this AGN we consider
that the CCF centroid, $\tau_{cent}$, is a more stable and representative
measure of the emission-line lags. The centroids quoted in Table~6 were
calculated at a level of 0.7 times the maximum CCF correlation
coefficient, $r_{max}$. Adopting a level of $0.8\,r_{max}$, as used in
previous papers (e.g., Wanders \etal 1997), usually gave a much larger
uncertainty on $\tau_{cent}$ due to the uncertainty on the value and
location of $r_{max}$.

There is no clear consensus on how best to assign uncertainties to CCF
parameters. Previous procedures have included comparison of different CCF
methods and analytical estimates (see Gaskell 1994 and references
therein). Here we have adopted a Monte-Carlo approach, in which large
numbers of surrogate data generated from the original data were
cross-correlated using the ICCF. The distribution of centroids calculated
from these surrogate data ICCFs --- the cross-correlation probability
distribution (CCPD, Maoz \& Netzer 1989) --- then gives an estimate of
the uncertainties (Peterson \etal 1998). The surrogate data were
generated in two ways.
\begin{enumerate}
\item Randomly generated, normally distributed
errors (Gaussian-deviates) derived from the observed flux uncertainties
were added to the observed fluxes. These data were then cross-correlated.
This is the ``flux randomization'' (FR) method described in Peterson \etal
(1997).

\item Gaussian-deviates were added {\em and} data points were randomly
excluded before cross-correlation. This is the FR/RSS method, where the
second step involves ``random subset selection'' (RSS) (Peterson \etal
1997), which gives an estimate of the effect of ``errant'' data
which may unduly influence the CCF. 
\end{enumerate}
Peterson \etal (1997) find that CCPDs calculated using the FR method
alone tend to underestimate the true uncertainty, whereas the FR/RSS
method is conservative in that it excludes real data.

We constructed CCPDs for 1000 runs of both the FR and FR/RSS methods and
from these calculated the mean centroids and standard deviations. These
are given in Table~6 as $\tau_{FR} \pm \sigma_{FR}$ and $\tau_{FR/RSS}
\pm \sigma_{FR/RSS}$. For some of the weaker emission lines and for
\Lya\,(Blue) it was necessary to exclude some outliers ($\approx 10$\% of
the runs) which lie in ``alias-peaks,'' spaced at approximately $\tau = 0 
\pm 160$ days, caused by features spaced at those intervals in the
second half of the continuum light-curves. These alias peaks are not seen
in either the \CIVnw\ or the other \Lya\ CCPDs. The data for lines where
aliasing was a particular problem, leading to greater uncertainty in
their true centroids, are bracketed in Table~6. We note that using a
standard deviation to describe the uncertainty is only appropriate if the
CCPD is Gaussian in shape, which is the case here. Gaskell (1988) and
Maoz \& Netzer (1989) suggest taking the range of lags within which 68\%
of the centroids contributing to the CCPDs occur. The median difference
between half the ranges calculated using that procedure and the $\sigma$
values given in Table~6 is only 1.5 days.

\subsubsection{The UV Continuum}

The sampling window auto-correlation function (ACF) shown in the top-left
window in Fig.~7 is clearly narrower than the continuum ACF. This
illustrates that the chosen sampling intervals were adequate to follow
most of the power in the continuum light-curves. The mean sampling
interval of 7.2 days limits the detection of lags on small timescales
between \cfs\ and \cfl , but the cross-correlation results are consistent
with no lag.

\subsubsection{The UV Emission Lines}

The \Lya\ and \CIVnw\ emission-line light-curves do not show the shortest
timescale features seen in the continuum light-curves. This suggests a
fairly large lag, as confirmed by the cross-correlation analysis. The CCF
amplitude remains high at large lags due to the linear trends in the
light-curves. Removing these trends reduces the CCF amplitudes but
does not significantly alter the values of $\tau_{cent}$.

The lags are larger for \Lya\ than for \CIVnw . We find no evidence for a
significant lag between the blue and red wings of \CIVnw , and the line
center also shows no significant lag relative to the wings. The \Lya\
data are consistent with these conclusions, although more uncertain. The
\CIVnw\ (Far-red) component can be used to further search for
velocity-dependent lags. The velocity limits for this component were
chosen to cover the same velocity interval ($\approx 6000$--11,000 km
s$^{-1}$) relative to the \CIVnw\ line-center as does the \NV\
measurement relative to \Lya\ (Section~3.1). This velocity region
overlaps with \HeII\ but the shape of the rms spectrum strongly suggests
the feature is not associated with \HeII. A cross-correlation analysis on
the resultant \CIVnw\,(Far-red) light-curve (Fig.~4) shows a smaller lag
than for the rest of \CIVnw . Repeating this procedure in the blue wing
of \CIVnw\ produces no significant difference in lag. Unfortunately, the
possible contaminating effect of \NV\ confuses the issue for \Lya .

\section{Discussion}

The UV data for 3C~390.3 clearly show that both the continuum and broad
emission-lines are highly variable. The detailed variability
characteristics show some similarities and some contrasts to those of the
intensively monitored Seyfert~1 galaxies.

The amplitude of the continuum variability is slightly lower in the
longer wavelength UV waveband according to the $F_{var}$ parameter, but the
difference is smaller than that seen in some Seyfert~1 galaxies (e.g., Clavel
\etal 1991; Crenshaw \etal 1996; Wanders \etal 1997). Over
the same nine month period in 1995, $F_{var} \approx 0.3$ for both the UV
and X-ray continua (Leighly \etal in preparation), suggesting that the
similarity in variability amplitude may apply to the entire ionizing
continuum. Such a similarity was not seen in NGC~4151, for which the
X-ray flux at $\sim 1.5$ keV varied by a factor of three more than the
1330\,\AA\ flux (Edelson \etal 1996). In contrast, the optical continuum
in 3C~390.3 appears to vary with a smaller amplitude than the UV
(Dietrich \etal 1998), behavior consistent with that seen in some
Seyfert~1 galaxies. In all of these AGN this trend may be partly due to
dilution of the non-stellar continuum by starlight from the host galaxy.

The broad UV emission-line variability correlates well with that of the
UV continuum in 3C~390.3, but the emission-line lags and the \CIVnw\
variability amplitude are larger than seen in Seyfert~1 galaxies of
comparable UV, but lower X-ray, luminosity. For example, in NGC~5548 the
\Lya\ lag is $\sim 10$ days (Clavel \etal 1991; Korista \etal 1995). The
results for 3C~390.3 are consistent with the trend seen in Seyfert~1
galaxies for the higher--ionization UV lines to vary on faster
timescales. However, the distinction in behaviour between what have been
commonly referred to as ``low'' and ``high'' ionization lines (Section~1)
appears more complex in 3C~390.3. The derived lags for the Balmer lines
($\sim 20\pm8$ days, Dietrich \etal 1998) are smaller than those for
both \Lya\ and \CIVnw , a trend not seen in Seyfert~1 galaxies. Whether these
contrasting characteristics are due to differences in the physical
conditions in the BLR or are a consequence of differences in the shape
and variability of the ionizing continuum is beyond the scope of this
paper. 

The difference in lag between 3C~390.3 and Seyfert~1 galaxies of
comparable luminosity was noted by Wamsteker \etal (1997) based on IUE
archival data, although the order-of-magnitude smaller mean sampling
interval used here makes the result more secure. Wamsteker \etal found
larger lags than those derived here, although even those may be
underestimates. Applying the same analysis techniques used here to the
measurements given by Wamsteker \etal we derive lags of $143 \pm 55$ and
$116 \pm 60$ days for \Lya\ and \CIVnw\ respectively compared to the lags
of $60\pm 24$ and $37\pm 14$ derived for the new data. Possible
variations in lag have been found previously for \Lya\ in the Seyfert~1
galaxy Fairall~9 (Rodr\'{\i}guez-Pascual 1997) and for \hb\ in NGC~5548
(Peterson \etal 1994). As for these AGN, the suggested variation in lag
in 3C~390.3 could be due to a change in the central ionizing luminosity
(e.g., O'Brien, Goad, \& Gondhalekar 1995), although this seems unlikely
since the mean UV continuum flux is only 30\% lower for the new campaign.
Other possibilities are that the physical properties of the emitting gas
have changed, for which there is direct evidence (see below), or that the
sampling rate of the archival data is insufficient to provide reliable
lags.

The values of $F_{var}$ found here for the continuum, all of the \CIVnw\
components, and \Lya\,(Red) are in good agreement with those derived by
Wamsteker \etal (1997). In contrast, $F_{var}$ for \Lya\,(Blue) and
\Lya\,(Center) are factors of 2 and 1.5 smaller (respectively) than for
the archival data. As discussed in Section 3.2, the change in the
\CIVnw /\Lya\ ratio between the new and archival data is striking,
particularly in the blue wing. The lower mean \Lya\ flux is partly due to
the lower mean continuum flux relative to the archival data, but even
allowing for that the efficiency of production of \Lya\ photons in the
new campaign is lower relative to \CIVnw . Overall, the new data indicate
a velocity-dependent change in the line emission, with \Lya\ affected
more than \CIVnw . 

Velocity-dependent changes in the UV line profiles in 3C~390.3 were noted
previously by Zheng (1996) and Wamsteker \etal (1997). Zheng found that
although the \Lya\ and \CIVnw\ blue wings were usually stronger than the
red in the IUE archival data, there were some very rare epochs when the
opposite was true. During the new campaign the blue wing was always
strongest. There is growing evidence for long-term variability in the BLR
gas properties and/or distribution. This evidence mainly comes from
long-term studies of Balmer-line variability in Seyfert~1 galaxies
(Wanders 1997), but there is also some evidence from UV data (e.g., in
NGC5548; Goad \& Koratkar 1998).

We find no significant lag between the blue and red wings of the \Lya\
and \CIVnw\ lines in either the new or archival data (adopting the larger
uncertainties calculated for the archival data using the FR/RSS method).
The new Balmer line data also show no evidence for radial motion
(Dietrich \etal 1998). These results imply that radial motion does not
dominate the kinematics of the UV/optical line-emitting gas in 3C~390.3.
The lack of a lag between the blue and red wings is broadly consistent
with the predictions of an accretion disk model. Detailed modelling is
required to determine if the different relative responses of the blue and
red wings of the \Lya\ and \CIVnw\ lines, at least at some epochs, the
multiple-components in the Balmer lines (Dietrich \etal ), and the
\CIVnw\ (Far-red) and possibly related Balmer line components can be
explained by such a model.

\section{Summary}

As part of an extensive multi-wavelength monitoring campaign, the
broad-line radio galaxy 3C~390.3 was observed every 6--10 days during the
period 1994 December 31 to 1996 March 5. Large amplitude continuum and
broad emission-line variability was detected. The principal conclusions
are as follows:

\begin{enumerate}

\item The UV continuum varied by a factor of seven, with little
difference in amplitude between 1370\AA\ and 1855\AA . During 1995, both
the shape and amplitude of the UV continuum light-curve are very similar
to that of the X-ray light-curve observed by ROSAT. This is in contrast
to the previously monitored Seyfert~1 galaxies in which the X-ray
amplitude is usually significantly larger than the UV. During the same
period, the optical continuum variability amplitude in 3C~390.3 was
smaller than that of the UV; behavior similar to that seen in Seyfert~1
galaxies.

\item The broad emission lines show correlated variability with that of
the UV continuum, with a general increase in flux through the campaign. 
The \Lya\ and \CIVnw\ emission-line variations lag the
continuum by 35--70 days. Overall, the emission-line lags are
systematically larger than those derived for Seyfert~1 galaxies of
comparable UV luminosity.

\item The trend in lag seen for the higher--ionization UV lines is consistent
with the idea of BLR ionization stratification found for Seyfert~1
galaxies, although the uncertainties are greater for the weaker lines in
3C~390.3. However, the \Lya\ and \CIVnw\ lags are significantly larger
than those found for the Balmer lines, in contrast to the results found
for Seyfert~1 galaxies, blurring the usual distinction between ``high'' and
``low'' ionization lines.

\item No evidence is found for a lag between the red and blue wings of
\CIVnw\ or \Lya , implying that radial motion does not dominate the
kinematics of the high-ionization line-emitting BLR gas. 

\item Comparing the new IUE data with archival data, there is some evidence
that the BLR in 3C~390.3 evolves. The relative strength of the blue and
red wings in \Lya\ and \CIV\ varies with time although the double-peaked
structure, seen in both the UV and optical line profiles, remains.

\acknowledgements{
We thank the staff at the VILSPA and Goddard IUE observatories for their
outstanding work on our behalf. We also gratefully acknowledge the
support of many colleagues, including those on the IUE time allocation
committees. This work has been supported in part by grant SFB328D
(Landessternwarte Heidelberg), NASA grants NAG5--3233 and NAG5--3497
to Ohio State University, and grant 97-02-17625 of the Russian Basic
Research Foundation.
}

\end{enumerate}

\newpage

\figcaption{
Average SWP spectrum of 3C~390.3 (top thick line) and rms
spectrum (lower thin line). The integration limits used for measuring the
continuum and major emission-line fluxes are shown below the rms spectrum
and are described in the text.}

\figcaption{ Continuum light-curves.}

\figcaption{ Light-curves for (a) \Lya\ and (b) \CIV .}

\figcaption{ Light-curves for the weaker UV lines and C\,{\sc iv}\,(Far-red).}

\figcaption{ Correlation between \cfs\ and \cfl\ (in units of $10^{-15}$ erg
cm$^{-2}$ s$^{-1}$ \AA$^{-1}$). The solid line is a straight-line fit
allowing for errors on both variables.}

\figcaption{ Correlation between the \CIVnw/\Lya\ ratio for the blue and red
wings as a function of line-wing flux and \cfs.}

\figcaption{ The continuum and emission-line ICCFs (solid lines) and DCFs
(error-bars). The top-left box shows the \cfs\ ACF and the sampling
window ACF. The other light-curves were all cross-correlated with that of
\cfs.}

\end{document}


\begin{deluxetable}{clcccc}
\tablenum{1}
\tablewidth{0pt}
\tablecaption{Observation Log}
\tablehead{
\colhead{SWP} &
\colhead{UT} &
\colhead{UT Start} &   
\colhead{JD Mid-point} &         
\colhead{Exp.\ Time} &  
\colhead{Note} \nl
\colhead{Image No.} &   
\colhead{Date } &
\colhead{} &   
\colhead{$-2440000$} & 
\colhead{(minutes)}  &   
\colhead{ } \nl
\colhead{(1)} & \colhead{(2)} & \colhead{(3)} & 
\colhead{(4)} & \colhead{(5)} & \colhead{(6)}
}
\startdata
53236 & 1994 Dec 31 & 20.7311 & 9718.364  & 410 & \nl
53286 & 1995 Jan 5 & 19.3281 & 9723.305  & 420 & \nl
53325 & 1995 Jan 11 & 16.2631 & 9729.178  & 420 & \nl
53465 & 1995 Jan 17 & 16.4417 & 9735.185  & 410 & \nl
53651 & 1995 Jan 24 & 08.9415 & 9741.873  & 315 & \nl
53825 & 1995 Feb 4 & 15.3592 & 9753.140  & 300 & 1\nl
53890 & 1995 Feb 11 & 18.1263 & 9760.255  & 265 & \nl
53928 & 1996 Feb 16 & 09.7267 & 9764.905  & 310 & \nl
53950 & 1995 Feb 22 & 09.2536 & 9770.886  & 355 & \nl
53996 & 1995 Feb 28 & 09.2804 & 9776.887  & 285 & \nl
54049 & 1995 Mar 6 & 07.3674 & 9782.807  & 343 & \nl
54108 & 1995 Mar 12 & 15.5560 & 9789.148  & 305 & \nl
54170 & 1995 Mar 18 & 15.5179 & 9795.146  & 330 & \nl
54217 & 1995 Mar 24 & 15.5671 & 9801.148  & 340 & \nl
54260 & 1995 Mar 30 & 15.2450 & 9807.135  & 345 & \nl
54300 & 1995 Apr 5 & 13.5896 & 9813.066  & 340 & \nl
54343 & 1995 Apr 9 & 13.6383 & 9817.068  & 335 & \nl
54453 & 1995 Apr 17 & 13.2365 & 9825.052  & 365 & \nl
54488 & 1995 Apr 23 & 03.6946 & 9830.654  & 140 & 2\nl
54489 & 1995 Apr 23 & 07.3015 & 9830.804  & 155 & 2\nl
54506 & 1995 Apr 26 & 05.4819 & 9833.729  & 330 & \nl
54536 & 1995 Apr 29 & 05.5161 & 9836.729  & 332 & \nl
54605 & 1995 May 5 & 03.6367 & 9842.651  & 318 & \nl
54655 & 1995 May 11 & 03.4326 & 9848.643  & 333 & \nl
54691 & 1995 May 17 & 03.6199 & 9854.650  & 325 & \nl
54737 & 1995 May 23 & 12.3044 & 9861.013  & 270 & \nl
54789 & 1995 May 29 & 11.7160 & 9866.988  & 260 & \nl
54857 & 1995 Jun 4 & 09.3038 & 9872.888  & 360 & \nl
54972 & 1995 Jun 10 & 09.1664 & 9878.882  & 345 & \nl
55008 & 1995 Jun 15 & 09.5492 & 9883.897  & 345 & \nl
55058 & 1995 Jun 22 & 09.5122 & 9890.896  & 330 & \nl
55162 & 1995 Jun 29 & 09.2940 & 9897.888  & 372 & \nl
55193 & 1995 Jul 3 & 07.3666 & 9901.807  & 360 & \nl
55236 & 1995 Jul 9 & 22.7790 & 9908.449  & 278 & \nl
55295 & 1995 Jul 15 & 22.5633 & 9914.440  & 290 & \nl
55339 & 1995 Jul 21 & 22.9281 & 9920.455  & 320 & \nl
55368 & 1995 Jul 27 & 22.7992 & 9926.450  & 337 & \nl
55408 & 1995 Aug 2 & 22.1089 & 9932.421  & 250 & \nl
55450 & 1995 Aug 8 & 21.7576 & 9938.406  & 260 & \nl
55532 & 1995 Aug 14 & 21.7831 & 9944.407  & 284 & \nl
55635 & 1995 Aug 21 & 21.6629 & 9951.402  & 315 & \nl
55749 & 1995 Aug 28 & 21.8886 & 9958.412  & 285 & \nl
55817 & 1995 Sep 4 & 19.5514 & 9965.314  & 305 & \nl
55908 & 1995 Sep 11 & 14.3917 & 9972.100  & 130 & 2 \nl
55909 & 1995 Sep 11 & 19.5281 & 9972.313  & 390 & \nl
55959 & 1995 Sep 18 & 17.3350 & 9979.222  & 641 & \nl
56016 & 1995 Sep 25 & 17.6154 & 9986.234  & 580 & 3 \nl
56036 & 1995 Sep 30 & 17.1876 & 9991.216  & 600 & 4 \nl
56059 & 1995 Oct 7 & 17.9060 & 9998.246  & 480 & 5 \nl
56087 & 1995 Oct 14 & 17.7842 & 10005.241\phn & 480 & 6 \nl
56113 & 1995 Oct 21 & 17.4051 & 10012.226\phn & 480 & 4 \nl
56130 & 1995 Oct 28 & 17.8189 & 10019.242\phn & 480 & \nl
56158 & 1995 Oct 7 & 15.4439 & 10029.144\phn & 480 & 7 \nl
56210 & 1995 Oct 11 & 15.5592 & 10039.148\phn & 480 & 8 \nl
56243 & 1995 Oct 27 & 15.3947 & 10049.142\phn & 480 & 4 \nl
56275 & 1995 Dec 7 & 17.9117 & 10059.246\phn & 440 & \nl
56304 & 1995 Dec 16 & 18.0831 & 10068.253\phn & 360 & 7 \nl
56358 & 1995 Dec 28 & 13.1900 & 10080.050\phn & 450 & 7 \nl
56385 & 1996 Jan 6 & 11.4000 & 10088.975\phn & 420 & 9 \nl
56582 & 1996 Jan 16 & 15.9649 & 10099.165\phn & 480 & 10 \nl
56637 & 1996 Jan 26 & 15.0621 & 10109.128\phn & 430 & 4 \nl
56759 & 1996 Feb 5 & 17.8221 & 10119.243\phn & 155 & 2 \nl
56810 & 1996 Feb 15 & 09.9738 & 10128.916\phn & 480 & 4 \nl
56851 & 1996 Feb 25 & 09.0300 & 10138.876\phn & 443 & 4 \nl
56893 & 1996 Mar 5 & 08.4969 & 10147.854\phn & 480 & 9 \nl
\tablecomments{(1)\ Image excluded because of unusally high background
and poor S/N.
(2)\ Image excluded because of under-exposure.
(3)\ 3 pixels saturated in \Lya.
(4)\ 1 pixel saturated in \Lya.
(5)\ 7 pixels saturated in \Lya.
(6)\ 5 pixels saturated in \Lya.
(7)\ 6 pixels saturated in \Lya.
(8)\ 10 pixels saturated in \Lya.
(9)\ 2 pixels saturated in \Lya.
(10)\ 8 pixels saturated in \Lya.}
\enddata
\end{deluxetable}

\newpage

\begin{deluxetable}{ccc}
\tablenum{2}
\tablewidth{0pt}
\tablecaption{Continuum Fluxes}
\tablehead{
\colhead{JD} &
\colhead{\cfs } &
\colhead{\cfl } \nl
\colhead{$-2440000$} &
\multicolumn{2}{c}{($10^{-15}$ erg cm$^{-2}$ s$^{-1}$ \AA$^{-1}$)} \nl
(1) & (2) & (3)
}
\startdata
  9718.364 &  2.76  $\pm$ 0.40 & 1.18  $\pm$ 0.27 \nl
  9723.306 &  3.57  $\pm$ 0.62 & 2.82  $\pm$ 0.31 \nl
  9729.178 &  3.21  $\pm$ 0.42 & 2.17  $\pm$ 0.34 \nl
  9735.186 &  1.86  $\pm$ 0.53 & 2.19  $\pm$ 0.38 \nl
  9741.873 &  3.30  $\pm$ 0.68 & 1.24  $\pm$ 0.45 \nl
  9760.255 &  2.76  $\pm$ 0.53 & 1.45  $\pm$ 0.47 \nl
  9764.905 &  3.48  $\pm$ 0.84 & 2.52  $\pm$ 0.89 \nl
  9770.886 &  2.47  $\pm$ 0.37 & 2.20  $\pm$ 0.41 \nl
  9776.887 &  2.60  $\pm$ 0.42 & 1.80  $\pm$ 0.50 \nl
  9782.807 &  3.75  $\pm$ 0.42 & 2.35  $\pm$ 0.33 \nl
  9789.148 &  4.74  $\pm$ 0.46 & 2.98  $\pm$ 0.38 \nl
  9795.146 &  7.44  $\pm$ 0.76 & 5.17  $\pm$ 0.50 \nl
  9801.148 &  5.62  $\pm$ 0.60 & 3.30  $\pm$ 0.44 \nl
  9807.135 &  4.25  $\pm$ 0.38 & 3.98  $\pm$ 0.43 \nl
  9813.066 &  5.51  $\pm$ 0.29 & 3.91  $\pm$ 0.31 \nl
  9817.068 &  4.20  $\pm$ 0.59 & 3.68  $\pm$ 0.50 \nl
  9825.052 &  4.68  $\pm$ 0.38 & 4.06  $\pm$ 0.39 \nl
  9833.729 &  5.15  $\pm$ 0.34 & 3.02  $\pm$ 0.30 \nl
  9836.729 &  4.44  $\pm$ 0.59 & 3.55  $\pm$ 0.57 \nl
  9842.651 &  5.16  $\pm$ 0.59 & 3.74  $\pm$ 0.46 \nl
  9848.643 &  4.34  $\pm$ 0.61 & 3.82  $\pm$ 0.50 \nl
  9854.650 &  4.83  $\pm$ 0.37 & 3.71  $\pm$ 0.32 \nl
  9861.013 &  5.50  $\pm$ 0.71 & 4.19  $\pm$ 0.46 \nl
  9866.988 &  4.41  $\pm$ 0.62 & 3.22  $\pm$ 0.51 \nl
  9872.888 &  4.81  $\pm$ 0.53 & 3.78  $\pm$ 0.38 \nl
  9878.882 &  5.12  $\pm$ 0.48 & 4.39  $\pm$ 0.55 \nl
  9883.897 &  4.77  $\pm$ 0.38 & 3.40  $\pm$ 0.33 \nl
  9890.896 &  5.21  $\pm$ 0.45 & 3.24  $\pm$ 0.36 \nl
  9897.888 &  5.05  $\pm$ 0.59 & 3.15  $\pm$ 0.42 \nl
  9901.807 &  4.04  $\pm$ 0.48 & 2.71  $\pm$ 0.36 \nl
  9908.449 &  3.57  $\pm$ 0.58 & 3.23  $\pm$ 0.44 \nl
  9914.440 &  3.33  $\pm$ 0.38 & 3.79  $\pm$ 0.37 \nl
  9920.455 &  3.99  $\pm$ 0.55 & 4.51  $\pm$ 0.43 \nl
  9926.450 &  4.82  $\pm$ 0.66 & 4.21  $\pm$ 0.37 \nl
  9932.421 &  5.10  $\pm$ 0.43 & 4.31  $\pm$ 0.37 \nl
  9938.406 &  5.69  $\pm$ 0.58 & 4.33  $\pm$ 0.48 \nl
  9944.407 &  6.64  $\pm$ 0.72 & 5.80  $\pm$ 0.80 \nl
  9951.402 &  7.73  $\pm$ 0.77 & 6.64  $\pm$ 0.57\tablenotemark{1} \nl
  9958.412 &  7.35  $\pm$ 0.51 & 5.11  $\pm$ 0.45 \nl
  9965.314 &  6.49  $\pm$ 0.60 & 5.55  $\pm$ 0.45 \nl
  9972.313 &  7.16  $\pm$ 0.45 & 5.27  $\pm$ 0.30 \nl
  9979.223 &  5.29  $\pm$ 0.50 & 4.20  $\pm$ 0.39 \nl
  9986.234 &  8.51  $\pm$ 0.49 & 5.70  $\pm$ 0.33 \nl
  9991.216 &  7.18  $\pm$ 0.41 & 5.32  $\pm$ 0.76 \nl
  9998.246 &  8.02  $\pm$ 1.02 & 5.34  $\pm$ 0.54 \nl
 10005.241\phn &  9.48  $\pm$ 1.24 & 5.38  $\pm$ 0.42 \nl
 10012.226\phn & 10.15  $\pm$ 0.51\phn & 6.48  $\pm$ 0.38 \nl
 10019.242\phn & 10.51  $\pm$ 0.31\phn & 7.42  $\pm$ 0.33 \nl
 10029.144\phn & 13.56  $\pm$ 0.47\phn & 8.18  $\pm$ 0.45 \nl
 10039.148\phn & 13.63  $\pm$ 0.68\phn & 8.99  $\pm$ 0.45\tablenotemark{2} \nl
 10049.142\phn & 12.95  $\pm$ 0.45\phn & 7.44  $\pm$ 0.31 \nl
 10059.246\phn &  9.37  $\pm$ 0.48 & 6.37  $\pm$ 0.38 \nl
 10068.253\phn &  9.43  $\pm$ 0.44 & 7.52  $\pm$ 0.50 \nl
 10080.050\phn &  9.32  $\pm$ 0.59 & 6.37  $\pm$ 0.56 \nl
 10088.975\phn &  8.91  $\pm$ 0.55 & 6.88  $\pm$ 0.39 \nl
 10099.165\phn &  8.49  $\pm$ 0.61 & 6.19  $\pm$ 0.40 \nl
 10109.128\phn &  9.36  $\pm$ 0.55 & 7.32  $\pm$ 0.40 \nl
 10128.916\phn & 10.84  $\pm$ 0.53\phn & 7.36  $\pm$ 0.37 \nl
 10138.876\phn & 12.03  $\pm$ 0.42\phn & 7.67  $\pm$ 0.27 \nl
 10147.854\phn & 12.05  $\pm$ 0.51\phn & 8.67  $\pm$ 0.35\tablenotemark{1} \nl
\enddata
\tablenotetext{1}{Waveband truncated to 1800 -- 1840~\AA.}
\tablenotetext{2}{Waveband truncated to 1800 -- 1870~\AA.}
\end{deluxetable}

\newpage

\begin{deluxetable}{ccccccccc}
\tablenum{3}
\tablewidth{0pt}
\tablecaption{Emission-line Fluxes for \Lya\ and \CIV }
\tablehead{
\colhead{JD} &
\multicolumn{4}{c}{\Lya } & \multicolumn{3}{c}{\CIV} \nl
$-2440000$ & Total & Blue & Center & Red & Total & Blue & Center & Red \nl
& \multicolumn{8}{c}{($10^{-13}$ erg cm$^{-2}$ s$^{-1}$)} \nl
(1) & (2) & (3) & (4) & (5) & (6) & (7) & (8) & (9)
}
\startdata
 9718.364      & \phn9.08$\pm$0.80     &  2.04$\pm$0.36  & 5.91$\pm$0.39  & 1.84$\pm$0.25  & 4.37$\pm$0.57       & 1.44$\pm$0.29  & 2.37$\pm$0.30  & 0.70$\pm$0.20 \nl
 9723.306      & 11.30$\pm$1.23        &  2.91$\pm$0.53  & 6.23$\pm$0.62  & 3.01$\pm$0.37  & 5.68$\pm$0.84       & 1.96$\pm$0.41  & 3.11$\pm$0.42  & 0.87$\pm$0.29 \nl
 9729.178      & \phn8.36$\pm$0.95     &  2.26$\pm$0.39  & 4.36$\pm$0.56  & 2.37$\pm$0.28  & 6.54$\pm$0.60       & 2.22$\pm$0.30  & 3.53$\pm$0.31  & 1.11$\pm$0.21 \nl
 9735.186      & 11.17$\pm$1.07        &  2.78$\pm$0.45  & 6.66$\pm$0.56  & 2.52$\pm$0.32  & 6.30$\pm$0.74       & 2.47$\pm$0.37  & 2.96$\pm$0.38  & 1.09$\pm$0.25 \nl
 9741.873      & \phn9.66$\pm$1.27     &  2.75$\pm$0.56  & 5.68$\pm$0.58  & 1.89$\pm$0.39  & 5.82$\pm$0.92       & 2.30$\pm$0.45  & 2.73$\pm$0.47  & 0.87$\pm$0.32 \nl
 9760.255      & \phn9.48$\pm$1.01     &  2.66$\pm$0.45  & 5.65$\pm$0.47  & 1.75$\pm$0.31  & 6.87$\pm$0.74       & 2.59$\pm$0.37  & 3.08$\pm$0.39  & 1.47$\pm$0.26 \nl
 9764.905      & 11.38$\pm$1.61        &  3.43$\pm$0.73  & 6.70$\pm$0.73  & 1.95$\pm$0.50  & 7.20$\pm$1.17       & 2.48$\pm$0.59  & 3.68$\pm$0.60  & 1.31$\pm$0.41 \nl
 9770.886      & 11.50$\pm$0.76        &  3.40$\pm$0.36  & 6.37$\pm$0.37  & 2.44$\pm$0.24  & 6.41$\pm$0.55       & 2.39$\pm$0.28  & 3.15$\pm$0.29  & 1.13$\pm$0.20 \nl
 9776.887      & 10.28$\pm$0.90        &  2.53$\pm$0.39  & 6.03$\pm$0.48  & 2.44$\pm$0.27  & 7.22$\pm$0.64       & 3.00$\pm$0.33  & 3.28$\pm$0.35  & 1.25$\pm$0.23 \nl
 9782.807      & 10.41$\pm$0.89        &  2.78$\pm$0.39  & 6.04$\pm$0.47  & 2.29$\pm$0.27  & 5.67$\pm$0.61       & 1.89$\pm$0.31  & 2.75$\pm$0.32  & 1.27$\pm$0.21 \nl
 9789.148      & 10.06$\pm$0.93        &  2.80$\pm$0.42  & 5.73$\pm$0.47  & 2.20$\pm$0.29  & 5.92$\pm$0.65       & 2.09$\pm$0.33  & 3.10$\pm$0.34  & 1.00$\pm$0.23 \nl
 9795.146      & 11.96$\pm$1.46        &  3.57$\pm$0.67  & 6.78$\pm$0.66  & 2.39$\pm$0.45  & 6.78$\pm$1.03       & 2.28$\pm$0.52  & 3.27$\pm$0.51  & 1.52$\pm$0.36 \nl
 9801.148      & 10.86$\pm$1.13        &  3.00$\pm$0.50  & 6.21$\pm$0.51  & 2.50$\pm$0.35  & 7.80$\pm$0.82       & 2.47$\pm$0.40  & 3.52$\pm$0.41  & 2.02$\pm$0.28 \nl
 9807.135      & 13.49$\pm$0.84        &  4.23$\pm$0.40  & 7.15$\pm$0.44  & 2.90$\pm$0.27  & 8.15$\pm$0.58       & 3.05$\pm$0.30  & 3.80$\pm$0.31  & 1.52$\pm$0.21 \nl
 9813.066      & 10.09$\pm$0.63        &  2.47$\pm$0.29  & 6.13$\pm$0.32  & 2.20$\pm$0.20  & 7.87$\pm$0.44       & 3.09$\pm$0.23  & 3.33$\pm$0.24  & 1.69$\pm$0.16 \nl
 9817.068      & 11.63$\pm$1.17        &  3.49$\pm$0.53  & 6.48$\pm$0.54  & 2.46$\pm$0.37  & 7.53$\pm$0.83       & 3.15$\pm$0.42  & 3.04$\pm$0.43  & 1.59$\pm$0.29 \nl
 9825.052      & 11.86$\pm$0.79        &  3.47$\pm$0.37  & 6.69$\pm$0.39  & 2.50$\pm$0.25  & 8.51$\pm$0.55       & 3.24$\pm$0.28  & 3.96$\pm$0.29  & 1.69$\pm$0.19 \nl
 9833.729      & 10.12$\pm$0.71        &  2.61$\pm$0.33  & 6.19$\pm$0.36  & 2.11$\pm$0.23  & 8.73$\pm$0.50       & 3.55$\pm$0.26  & 3.86$\pm$0.27  & 1.71$\pm$0.18 \nl
 9836.729      & 12.84$\pm$1.15        &  3.63$\pm$0.52  & 7.06$\pm$0.54  & 2.99$\pm$0.36  & 10.10$\pm$0.83\phn  & 3.79$\pm$0.42  & 4.24$\pm$0.43  & 2.46$\pm$0.29 \nl
 9842.651      & 12.32$\pm$1.16        &  3.71$\pm$0.51  & 6.78$\pm$0.56  & 2.65$\pm$0.36  & 9.28$\pm$0.81       & 4.17$\pm$0.41  & 3.35$\pm$0.42  & 2.09$\pm$0.28 \nl
 9848.643      & 12.60$\pm$1.24        &  4.17$\pm$0.55  & 6.67$\pm$0.63  & 2.61$\pm$0.38  & 9.28$\pm$0.86       & 3.98$\pm$0.43  & 3.73$\pm$0.46  & 1.98$\pm$0.29 \nl
 9854.650      & 10.97$\pm$0.76        &  3.01$\pm$0.35  & 6.18$\pm$0.37  & 2.49$\pm$0.24  & 7.71$\pm$0.54       & 3.44$\pm$0.28  & 3.16$\pm$0.28  & 1.47$\pm$0.19 \nl
 9861.013      & 11.94$\pm$1.34        &  3.31$\pm$0.60  & 6.60$\pm$0.60  & 2.79$\pm$0.41  & 8.78$\pm$0.97       & 3.40$\pm$0.48  & 3.50$\pm$0.49  & 2.11$\pm$0.33 \nl
 9866.988      & \phn9.17$\pm$1.17     &  2.87$\pm$0.53  & 5.14$\pm$0.51  & 1.81$\pm$0.37  & 7.82$\pm$0.87       & 3.25$\pm$0.44  & 3.02$\pm$0.44  & 1.96$\pm$0.30 \nl
 9872.888      & 12.12$\pm$1.08        &  3.48$\pm$0.48  & 6.60$\pm$0.54  & 2.75$\pm$0.34  & 7.93$\pm$0.74       & 3.05$\pm$0.37  & 3.67$\pm$0.38  & 1.51$\pm$0.25 \nl
 9878.882      & 12.75$\pm$1.00        &  4.16$\pm$0.46  & 6.73$\pm$0.51  & 2.70$\pm$0.31  & 8.44$\pm$0.67       & 3.54$\pm$0.34  & 3.32$\pm$0.34  & 1.89$\pm$0.23 \nl
 9883.897      & 11.20$\pm$0.78        &  3.27$\pm$0.36  & 6.23$\pm$0.39  & 2.38$\pm$0.25  & 10.01$\pm$0.56\phn  & 4.57$\pm$0.29  & 4.11$\pm$0.29  & 1.67$\pm$0.19 \nl
 9890.896      & 11.22$\pm$0.94        &  3.43$\pm$0.43  & 6.42$\pm$0.48  & 2.19$\pm$0.29  & 9.03$\pm$0.66       & 3.81$\pm$0.34  & 3.65$\pm$0.35  & 2.01$\pm$0.23 \nl
 9897.888      & 11.67$\pm$1.15        &  3.27$\pm$0.51  & 6.81$\pm$0.54  & 2.37$\pm$0.35  & 8.76$\pm$0.81       & 3.58$\pm$0.40  & 3.91$\pm$0.40  & 1.72$\pm$0.28 \nl
 9901.807      & 11.76$\pm$1.00        &  3.31$\pm$0.44  & 6.58$\pm$0.51  & 2.57$\pm$0.31  & 9.13$\pm$0.69       & 4.18$\pm$0.35  & 3.65$\pm$0.36  & 1.69$\pm$0.23 \nl
 9908.449      & 11.36$\pm$1.12        &  3.33$\pm$0.50  & 6.45$\pm$0.52  & 2.39$\pm$0.36  & 9.20$\pm$0.81       & 3.47$\pm$0.40  & 3.88$\pm$0.41  & 2.27$\pm$0.28 \nl
 9914.440      & 11.93$\pm$0.80        &  3.77$\pm$0.38  & 6.52$\pm$0.40  & 2.50$\pm$0.25  & 8.58$\pm$0.57       & 3.40$\pm$0.30  & 3.68$\pm$0.30  & 1.88$\pm$0.20 \nl
 9920.455      & 11.83$\pm$1.08        &  3.62$\pm$0.49  & 6.57$\pm$0.52  & 2.49$\pm$0.34  & 8.09$\pm$0.77       & 3.49$\pm$0.39  & 3.53$\pm$0.39  & 1.48$\pm$0.27 \nl
 9926.450      & 11.50$\pm$1.24        &  3.35$\pm$0.55  & 6.50$\pm$0.55  & 2.43$\pm$0.38  & 8.41$\pm$0.90       & 3.48$\pm$0.44  & 3.57$\pm$0.46  & 1.67$\pm$0.31 \nl
 9932.421      & 10.88$\pm$0.88        &  3.00$\pm$0.41  & 6.45$\pm$0.43  & 2.20$\pm$0.28  & 8.58$\pm$0.64       & 3.42$\pm$0.33  & 3.71$\pm$0.33  & 1.80$\pm$0.23 \nl
 9938.406      & 11.06$\pm$1.10        &  3.45$\pm$0.50  & 6.07$\pm$0.50  & 2.13$\pm$0.34  & 7.79$\pm$0.80       & 3.29$\pm$0.40  & 3.44$\pm$0.41  & 1.39$\pm$0.28 \nl
 9944.407      & 12.53$\pm$1.39        &  3.62$\pm$0.62  & 7.24$\pm$0.64  & 2.57$\pm$0.43  & 10.20$\pm$1.01\phn  & 4.02$\pm$0.51  & 4.42$\pm$0.51  & 2.34$\pm$0.35 \nl
 9951.402      & 13.13$\pm$1.45        &  4.25$\pm$0.64  & 7.18$\pm$0.66  & 2.55$\pm$0.44  & 10.84$\pm$1.05\phn  & 4.46$\pm$0.52  & 4.62$\pm$0.52  & 2.16$\pm$0.36 \nl
 9958.412      & 11.91$\pm$1.02        &  3.41$\pm$0.47  & 6.77$\pm$0.48  & 2.61$\pm$0.32  & 8.73$\pm$0.72       & 3.39$\pm$0.37  & 3.64$\pm$0.37  & 2.05$\pm$0.25 \nl
 9965.314      & 13.11$\pm$1.16        &  4.13$\pm$0.53  & 7.12$\pm$0.53  & 2.75$\pm$0.36  & 8.77$\pm$0.83       & 3.49$\pm$0.41  & 3.61$\pm$0.41  & 2.00$\pm$0.29 \nl
 9972.313      & 10.49$\pm$0.90        &  2.97$\pm$0.42  & 6.03$\pm$0.43  & 2.21$\pm$0.28  & 7.32$\pm$0.63       & 3.16$\pm$0.32  & 2.88$\pm$0.32  & 1.53$\pm$0.22 \nl
 9979.223      & \phn9.41$\pm$0.95     &  2.87$\pm$0.43  & 5.30$\pm$0.43  & 1.91$\pm$0.30  & 7.80$\pm$0.68       & 2.77$\pm$0.34  & 3.31$\pm$0.35  & 2.04$\pm$0.23 \nl
 9986.234      & 13.52$\pm$1.05        &  4.66$\pm$0.45  & 6.19$\pm$0.57  & 3.57$\pm$0.32  & 11.35$\pm$0.69\phn  & 4.79$\pm$0.35  & 4.46$\pm$0.37  & 2.62$\pm$0.24 \nl
 9991.216      & 12.43$\pm$0.85        &  3.68$\pm$0.38  & 6.69$\pm$0.43  & 2.83$\pm$0.26  & 9.68$\pm$0.58       & 3.90$\pm$0.29  & 3.91$\pm$0.29  & 2.27$\pm$0.20 \nl
 9998.246      & 11.90$\pm$2.08        &  3.61$\pm$0.98  & 5.70$\pm$1.03  & 3.22$\pm$0.60  & 11.29$\pm$1.38\phn  & 4.77$\pm$0.68  & 4.54$\pm$0.70  & 2.58$\pm$0.46 \nl
 10005.241\phn & \phn9.76$\pm$2.34     &  2.49$\pm$1.04  & 5.81$\pm$1.07  & 2.07$\pm$0.71  & 11.22$\pm$1.62\phn  & 4.44$\pm$0.78  & 4.91$\pm$0.79  & 2.46$\pm$0.53 \nl
 10012.226\phn & 12.19$\pm$1.05        &  3.51$\pm$0.47  & 6.66$\pm$0.53  & 3.01$\pm$0.33  & 11.24$\pm$0.72\phn  & 4.69$\pm$0.37  & 4.58$\pm$0.36  & 2.50$\pm$0.25 \nl
 10019.242\phn & 13.98$\pm$0.71        &  4.16$\pm$0.33  & 7.57$\pm$0.39  & 3.15$\pm$0.22  & 11.94$\pm$0.46\phn  & 5.01$\pm$0.24  & 5.02$\pm$0.25  & 2.42$\pm$0.16 \nl
 10029.144\phn & 13.83$\pm$1.04        &  4.09$\pm$0.51  & 7.01$\pm$0.55  & 3.60$\pm$0.31  & 12.30$\pm$0.66\phn  & 4.79$\pm$0.34  & 5.07$\pm$0.35  & 3.00$\pm$0.23 \nl
 10039.148\phn & 14.39$\pm$1.49        &  4.16$\pm$0.63  & 8.03$\pm$0.86  & 3.44$\pm$0.42  & 13.44$\pm$0.91\phn  & 5.49$\pm$0.45  & 5.69$\pm$0.45  & 2.98$\pm$0.30 \nl
 10049.142\phn & 14.46$\pm$0.95        &  3.34$\pm$0.43  & 8.28$\pm$0.49  & 4.06$\pm$0.29  & 14.35$\pm$0.63\phn  & 5.44$\pm$0.32  & 5.80$\pm$0.32  & 3.74$\pm$0.21 \nl
 10059.246\phn & 15.31$\pm$1.10        &  4.46$\pm$0.46  & 7.78$\pm$0.66  & 4.16$\pm$0.31  & 14.24$\pm$0.66\phn  & 5.37$\pm$0.33  & 5.68$\pm$0.34  & 3.87$\pm$0.23 \nl
 10068.253\phn & 15.42$\pm$1.00        &  4.50$\pm$0.44  & 8.06$\pm$0.58  & 4.12$\pm$0.31  & 14.22$\pm$0.61\phn  & 5.94$\pm$0.31  & 5.66$\pm$0.32  & 3.41$\pm$0.21 \nl
 10080.050\phn & 14.89$\pm$1.29        &  4.54$\pm$0.63  & 7.35$\pm$0.66  & 4.09$\pm$0.39  & 13.91$\pm$0.84\phn  & 5.59$\pm$0.43  & 5.47$\pm$0.44  & 3.45$\pm$0.29 \nl
 10088.975\phn & 17.24$\pm$1.18        &  5.32$\pm$0.49  & 9.09$\pm$0.67  & 4.02$\pm$0.36  & 14.57$\pm$0.75\phn  & 6.10$\pm$0.37  & 5.73$\pm$0.38  & 3.41$\pm$0.25 \nl
 10099.165\phn & 14.94$\pm$1.39        &  4.44$\pm$0.62  & 7.75$\pm$0.81  & 3.70$\pm$0.40  & 14.82$\pm$0.84\phn  & 6.20$\pm$0.43  & 5.60$\pm$0.43  & 3.67$\pm$0.28 \nl
 10109.128\phn & 16.22$\pm$1.14        &  4.58$\pm$0.49  & 8.66$\pm$0.61  & 4.15$\pm$0.34  & 12.94$\pm$0.75\phn  & 5.07$\pm$0.37  & 5.46$\pm$0.37  & 3.02$\pm$0.25 \nl
 10128.916\phn & 16.50$\pm$1.08        &  5.16$\pm$0.50  & 8.44$\pm$0.53  & 4.11$\pm$0.34  & 14.85$\pm$0.75\phn  & 5.85$\pm$0.38  & 6.29$\pm$0.39  & 3.48$\pm$0.25 \nl
 10138.876\phn & 16.16$\pm$0.91        &  4.32$\pm$0.42  & 8.22$\pm$0.48  & 4.75$\pm$0.29  & 17.34$\pm$0.61\phn  & 6.67$\pm$0.31  & 7.69$\pm$0.33  & 3.90$\pm$0.21 \nl
 10147.854\phn & 17.62$\pm$1.06        &  5.20$\pm$0.47  & 9.05$\pm$0.55  & 4.68$\pm$0.33  & 15.68$\pm$0.71\phn  & 6.05$\pm$0.36  & 5.85$\pm$0.36  & 4.51$\pm$0.25 \nl
\enddata
\end{deluxetable}

\newpage

\begin{deluxetable}{ccccc}
\tablenum{4}
\tablewidth{0pt}
\tablecaption{Emission-line Fluxes for \NV , \SiIV , \HeII , 
and C\,{\sc iv}\,(far-red)}
\tablehead{
\colhead{JD} &
\NV &
\SiIV &
\HeII &
C\,{\sc iv}\,(Far-red) \nl
$-2440000$) & 
\multicolumn{4}{c}{($10^{-13}$ erg cm$^{-2}$ s$^{-1}$)} \nl
(1) & (2) & (3) & (4) & (5) 
}
\startdata
 9718.364      & 0.42$\pm$0.15  & 0.87$\pm$0.37   & 1.04$\pm$0.61  & 0.01$\pm$0.33 \nl
 9723.306      & 1.41$\pm$0.22  & 0.45$\pm$0.55   & 0.40$\pm$0.91  & 0.46$\pm$0.49 \nl
 9729.178      & 0.99$\pm$0.16  & 0.66$\pm$0.39   & 0.59$\pm$0.64  & 0.29$\pm$0.35 \nl
 9735.186      & 1.21$\pm$0.19  & 0.11$\pm$0.47   & 0.37$\pm$0.79  & 0.50$\pm$0.43 \nl
 9741.873      & 0.80$\pm$0.24  & 0.22$\pm$0.59   & 1.80$\pm$1.00  & 0.65$\pm$0.54 \nl
 9760.255      & 0.73$\pm$0.19  & 0.30$\pm$0.47   & 2.33$\pm$0.79  & 0.93$\pm$0.44 \nl
 9764.905      & 0.55$\pm$0.32  & 0.97$\pm$0.77   & 2.61$\pm$1.26  & 0.86$\pm$0.69 \nl
 9770.886      & 1.01$\pm$0.15  & 0.88$\pm$0.35   & 3.00$\pm$0.57  & 0.90$\pm$0.33 \nl
 9776.887      & 1.12$\pm$0.17  & 0.45$\pm$0.41   & 1.77$\pm$0.66  & 0.77$\pm$0.39 \nl
 9782.807      & 0.65$\pm$0.16  & 0.21$\pm$0.40   & 3.00$\pm$0.65  & 0.98$\pm$0.37 \nl
 9789.148      & 0.82$\pm$0.17  & 1.08$\pm$0.43   & 2.58$\pm$0.70  & 0.46$\pm$0.39 \nl
 9795.146      & 1.14$\pm$0.28  & 0.99$\pm$0.68   & 1.89$\pm$1.12  & 1.10$\pm$0.61 \nl
 9801.148      & 1.09$\pm$0.22  & 0.54$\pm$0.53   & 4.41$\pm$0.89  & 1.94$\pm$0.48 \nl
 9807.135      & 1.30$\pm$0.16  & 1.22$\pm$0.38   & 2.80$\pm$0.60  & 1.54$\pm$0.36 \nl
 9813.066      & 1.04$\pm$0.12  & 0.70$\pm$0.29   & 2.03$\pm$0.45  & 1.48$\pm$0.27 \nl
 9817.068      & 1.08$\pm$0.23  & 1.39$\pm$0.55   & 2.81$\pm$0.89  & 1.19$\pm$0.50 \nl
 9825.052      & 1.06$\pm$0.15  & 0.26$\pm$0.36   & 1.85$\pm$0.58  & 1.38$\pm$0.33 \nl
 9833.729      & 0.76$\pm$0.14  & 1.43$\pm$0.33   & 4.03$\pm$0.53  & 1.34$\pm$0.31 \nl
 9836.729      & 1.24$\pm$0.22  & 1.20$\pm$0.54   & 3.37$\pm$0.89  & 1.71$\pm$0.49 \nl
 9842.651      & 1.07$\pm$0.22  & 0.85$\pm$0.52   & 2.07$\pm$0.87  & 0.83$\pm$0.48 \nl
 9848.643      & 1.27$\pm$0.23  & 2.10$\pm$0.55   & 3.16$\pm$0.91  & 1.48$\pm$0.50 \nl
 9854.650      & 1.22$\pm$0.15  & 0.78$\pm$0.35   & 0.57$\pm$0.56  & 0.63$\pm$0.32 \nl
 9861.013      & 1.05$\pm$0.26  & 1.03$\pm$0.62   & 3.74$\pm$1.05  & 1.89$\pm$0.58 \nl
 9866.988      & 0.71$\pm$0.23  & 2.46$\pm$0.56   & 2.24$\pm$0.93  & 1.27$\pm$0.52 \nl
 9872.888      & 1.20$\pm$0.21  & 0.78$\pm$0.47   & 2.56$\pm$0.79  & 1.27$\pm$0.43 \nl
 9878.882      & 1.18$\pm$0.19  & 1.32$\pm$0.44   & 1.70$\pm$0.71  & 0.97$\pm$0.39 \nl
 9883.897      & 0.92$\pm$0.15  & 0.15$\pm$0.36   & 1.70$\pm$0.58  & 0.99$\pm$0.33 \nl
 9890.896      & 0.50$\pm$0.17  & 0.37$\pm$0.42   & 2.71$\pm$0.68  & 1.13$\pm$0.38 \nl
 9897.888      & 0.96$\pm$0.21  & 1.44$\pm$0.53   & 2.47$\pm$0.88  & 1.02$\pm$0.47 \nl
 9901.807      & 0.98$\pm$0.18  & 0.75$\pm$0.44   & 2.43$\pm$0.72  & 0.94$\pm$0.40 \nl
 9908.449      & 1.12$\pm$0.21  & 1.94$\pm$0.53   & 2.90$\pm$0.86  & 1.27$\pm$0.48 \nl
 9914.440      & 1.02$\pm$0.15  & 1.35$\pm$0.36   & 2.14$\pm$0.60  & 0.65$\pm$0.34 \nl
 9920.455      & 0.97$\pm$0.20  & 1.33$\pm$0.50   & 1.22$\pm$0.82  & 0.69$\pm$0.45 \nl
 9926.450      & 0.74$\pm$0.23  & 0.84$\pm$0.58   & 1.93$\pm$0.98  & 1.25$\pm$0.53 \nl
 9932.421      & 0.75$\pm$0.17  & 1.26$\pm$0.41   & 2.32$\pm$0.67  & 1.07$\pm$0.38 \nl
 9938.406      & 0.99$\pm$0.21  & 1.34$\pm$0.51   & 2.26$\pm$0.86  & 1.00$\pm$0.47 \nl
 9944.407      & 1.03$\pm$0.26  & 1.13$\pm$0.65   & 2.35$\pm$1.08  & 1.46$\pm$0.60 \nl
 9951.402      & 1.09$\pm$0.27  & 0.62$\pm$0.68   & 0.51$\pm$1.13  & 1.08$\pm$0.62 \nl
 9958.412      & 1.18$\pm$0.19  & 1.73$\pm$0.48   & 2.82$\pm$0.77  & 1.18$\pm$0.43 \nl
 9965.314      & 1.39$\pm$0.22  & 0.82$\pm$0.53   & 2.02$\pm$0.89  & 1.32$\pm$0.49 \nl
 9972.313      & 1.14$\pm$0.17  & 1.14$\pm$0.42   & 1.86$\pm$0.67  & 1.11$\pm$0.37 \nl
 9979.223      & 1.04$\pm$0.18  & 1.22$\pm$0.45   & 2.20$\pm$0.74  & 1.42$\pm$0.40 \nl
 9986.234      & 1.75$\pm$0.19  & 1.80$\pm$0.44   & 3.26$\pm$0.72  & 2.02$\pm$0.40 \nl
 9991.216      & 1.31$\pm$0.16  & 1.78$\pm$0.38   & 3.13$\pm$0.61  & 1.76$\pm$0.34 \nl
 9998.246      & 1.31$\pm$0.36  & 2.64$\pm$0.89   & 3.28$\pm$1.49  & 1.85$\pm$0.80 \nl
 10005.241\phn & 0.94$\pm$0.43  & 1.24$\pm$1.05   & 3.62$\pm$1.78  & 1.87$\pm$0.93 \nl
 10012.226\phn & 1.43$\pm$0.20  & 1.67$\pm$0.48   & 2.72$\pm$0.76  & 1.68$\pm$0.41 \nl
 10019.242\phn & 1.52$\pm$0.13  & 1.71$\pm$0.30   & 3.36$\pm$0.47  & 1.98$\pm$0.27 \nl
 10029.144\phn & 1.72$\pm$0.19  & 1.79$\pm$0.43   & 4.08$\pm$0.70  & 2.62$\pm$0.39 \nl
 10039.148\phn & 1.63$\pm$0.24  & 2.73$\pm$0.59   & 4.12$\pm$0.99  & 2.44$\pm$0.52 \nl
 10049.142\phn & 1.90$\pm$0.18  & 1.14$\pm$0.41   & 5.95$\pm$0.67  & 2.92$\pm$0.37 \nl
 10059.246\phn & 2.09$\pm$0.18  & 1.31$\pm$0.43   & 5.04$\pm$0.70  & 3.00$\pm$0.39 \nl
 10068.253\phn & 1.95$\pm$0.17  & 1.84$\pm$0.40   & 2.76$\pm$0.65  & 2.16$\pm$0.35 \nl
 10080.050\phn & 1.60$\pm$0.23  & 2.28$\pm$0.54   & 3.37$\pm$0.88  & 1.98$\pm$0.48 \nl
 10088.975\phn & 1.82$\pm$0.20  & 2.40$\pm$0.49   & 3.12$\pm$0.80  & 1.78$\pm$0.43 \nl
 10099.165\phn & 1.80$\pm$0.23  & 2.07$\pm$0.54   & 4.65$\pm$0.89  & 2.61$\pm$0.48 \nl
 10109.128\phn & 2.02$\pm$0.21  & 1.91$\pm$0.49   & 3.01$\pm$0.80  & 1.67$\pm$0.43 \nl
 10128.916\phn & 1.82$\pm$0.21  & 1.98$\pm$0.49   & 5.20$\pm$0.79  & 2.45$\pm$0.43 \nl
 10138.876\phn & 1.80$\pm$0.17  & 1.90$\pm$0.39   & 4.34$\pm$0.62  & 2.36$\pm$0.35 \nl
 10147.854\phn & 2.01$\pm$0.20  & 1.54$\pm$0.46   & 4.22$\pm$0.76  & 3.33$\pm$0.42 \nl

\enddata
\end{deluxetable}

\clearpage

\newpage

\begin{deluxetable}{lcccc}
\tablenum{5}
\tablewidth{250pt}
\tablecaption{Variability Parameters}
\tablehead{
\colhead{Feature} &
\colhead{{$\overline{F}$}\tablenotemark{a}} &
\colhead{{$\sigma_F$}\tablenotemark{a}} &
\colhead{$R_{max}$} &
\colhead{$F_{var}$} \nl
\colhead{(1)} & (2) & (3) & (4) & (5) 
}
\startdata
\cfs & 6.33 & 2.99 & 7.33 & 0.46 \nl
\cfl & 4.57 & 1.95 & 7.62 & 0.41 \nl
\nl
\Lya & 12.22\phn & 2.10 & 2.11 & 0.14 \nl
\Lya\,(Blue) & 3.55 & 0.74 & 2.61 & 0.15 \nl
\Lya\,(Center) & 6.72 & 0.93 & 2.08 & 0.10 \nl
\Lya\,(Red) & 2.79 & 0.75 & 2.71 & 0.24 \nl
\nl
\CIV & 9.56 & 2.92 & 3.97 & 0.29 \nl
\CIVnw\,(Blue) & 3.81 & 1.24 & 4.63 & 0.31 \nl
\CIVnw\,(Center) & 4.06 & 1.07 & 3.24 & 0.25 \nl
\CIVnw\,(Red) & 2.11 & 0.86 & 6.44 & 0.39 \nl
\nl
\NV & 1.21 & 0.41 & 4.97 & 0.29 \nl
\SiIV & 1.24 & 0.65 & (24.82)\phn & 0.33 \nl
\CIVnw\,(Far-red) & 1.42 & 0.70 & (330.00)\phn\phn & 0.37 \nl
\HeII & 2.70 & 1.19 & (16.08)\phn & 0.31 \nl
\enddata
\tablenotetext{a}{Units are $10^{-15}$ erg cm$^{-2}$ s$^{-1}$ \AA$^{-1}$ for
continuum fluxes and $10^{-13}$ erg cm$^{-2}$ s$^{-1}$ for emission
lines.}
\end{deluxetable}

\newpage

\begin{deluxetable}{lcccccc}
\tablenum{6}
\tablewidth{0pt}
\tablecaption{Cross-Correlation Results}
\tablehead{
\colhead{Feature} &
\colhead{$r_{max}$} &
\colhead{FWHM} &
\colhead{$\tau_{peak}$} &
\colhead{$\tau_{cent}$} &
\colhead{$\tau_{FR}\pm\sigma_{FR}$} &
\colhead{$\tau_{FR/RSS}\pm\sigma_{FR/RSS}$} \nl
\colhead{\ } &
\colhead{\ } &
\colhead{(days)} &
\colhead{(days)} &
\colhead{(days)} &
\colhead{(days)} &
\colhead{(days)} \nl
\colhead{(1)} & (2) & (3) & (4) & (5) & (6) & (7)
}
\startdata
\cfs \ (ACF) & 1.00 & \phn47 & \phn0 & \phn0 & \nodata & \nodata \nl
\cfl \ (ACF) & 1.00 & \phn58 & \phn0 & \phn0 & \nodata & \nodata \nl
\cfl & 0.95 & \phn61 & \phn$-$2\phs & \phn3 & $-$10$\pm$11\phs & \phn$-$3$\pm$18\phs \nl
\nl
\Lya & 0.80 & 278 & 93 & 67 & 59$\pm$12 & 57$\pm$22\nl
\Lya\,(Blue) & 0.73 & \nodata & 56 & 62 & (61$\pm$25) & (66$\pm$35)\nl
\Lya\,(Center) & 0.78 & 214 & 83 & 65 & 63$\pm$18 & 63$\pm$26\nl
\Lya\,(Red) & 0.85 & \nodata & 26 & 54 & 54$\pm$9\phn & 51$\pm$17\nl
\nl
\CIV & 0.88 & \nodata & 15 & 41 & 40$\pm$7\phn & 36$\pm$14 \nl
\CIVnw\,(Blue) & 0.88 & \nodata & 45 & 41 & 34$\pm$9\phn & 35$\pm$14 \nl
\CIVnw\,(Center) & 0.85 & \nodata & \phn3 & 50 & 45$\pm$8\phn & 40$\pm$17 \nl
\CIVnw\,(Red) & 0.89 & \nodata & 16 & 35 & 36$\pm$9\phn & 34$\pm$16 \nl
\nl
\NV & 0.87 & \nodata & 15 & 43 & 36$\pm$13 & 34$\pm$17 \nl
\SiIV & (0.70) & \nodata & ($-$27)\phs & \phn(9) & \phn(5$\pm$27) & \phn(7$\pm$31) \nl
\CIVnw\,(Far-red) & 0.89 & 202 & \phn9 & 11 & 11$\pm$9\phn & 11$\pm$14 \nl
\HeII & (0.73) & (177) & \phn(4) & (12) & \phn(9$\pm$11) & (14$\pm$18) \nl
\enddata
\end{deluxetable}